\documentclass{elsart}
\usepackage{natbib}
\usepackage{epsfig}
\usepackage{unites2e}
\begin{document}
\runauthor{LeBohec and Holder}
\begin{frontmatter}
\title{The Cosmic Ray Background as a Tool for Relative Calibration
of Atmospheric Cherenkov Telescopes}

\author[MCSD]{S. LeBohec}
\author[Leeds]{and J. Holder}
\address[MCSD]{Department of Physics and Astronomy,ISU\\
               Ames, IA, 50011, USA}
\address[Leeds]{Department of Physics and Astronomy,\\ 
                University of Leeds, Leeds, UK}
      
\begin{abstract}

The atmosphere is an intrinsic part of any ground based Cherenkov $\gamma$-ray
telescope, and the telescope response is therefore sensitive to unpredictable
changes in the atmospheric transparency which are difficult to measure and
interpret in the absence of a calibrated beam of high energy $\gamma$-rays. In
this paper, we use the detector response to Cherenkov emission from cosmic ray
initiated air showers to obtain a relative calibration for data obtained under
different instrumental and atmospheric conditions as well as over a range of
source angles to the Zenith. We show that such a relative calibration is
useful and efficient for data selection, for correcting the measured
$\gamma$-ray rate and for inter-calibration between the elements of an array of
Cherenkov telescopes.

\vspace{1pc}
\end{abstract}
\end{frontmatter}

\section{Introduction}
Atmospheric Cherenkov detectors cannot be calibrated using a test beam and the
estimation of their sensitivity strongly depends on Monte Carlo simulation
programs in which are modeled the atmosphere and the various elements of the
detectors. Simulations usually assume a set of fixed conditions while the
overall efficiency of the experiment can vary in time due to a number of
factors. The most important cause of these variations is the atmosphere
itself, and measuring and modeling changes in the $\gamma$-ray detection
efficiency due to changing atmospheric conditions is complex. The slow
degradation of optical elements until their recoating or replacement or the
occasional readjustment of photo-detector gains also affects the sensitivity
of the experiment. When measuring the $\gamma$-ray flux from a source, one
must correct for these effects.

In this paper we present a method used for the Whipple Atmospheric Cherenkov
Imaging Telescope to estimate an overall relative efficiency factor. We also
validate the method using observations of the Crab Nebula and present some
applications. In its basic form the method is based on the analysis of data
taken toward the Zenith \cite{Mohanty} and this is presented first. We have
realized that the method can be generalized in a way which incorporates the
effects of the Zenith angle at which observations are made. While detailed
simulations will always be necessary in order to understand variations in
telescope sensitivity, a simple correction such as that presented here is a
useful tool which may be particularly important when studying the time
variability of $\gamma$-ray sources. An example of this is the case of flaring
active galactic nuclei (AGN) which may be observed over a long period of time
and a wide range of Zenith angles and atmospheric conditions. We also present
the application of a similar method to CELESTE, a Cherenkov wavefront sampling
experiment, which illustrates the utility of the technique as a way of
obtaining a relative calibration between individual elements of a detector
array.

\section{Relative calibration at fixed Zenith angle}
\subsection{The method}

Each recorded Cherenkov event can be characterized by its luminosity, $Q$, the
definition of which may depend on the specific experiment. A relative
throughput factor, $F$, between two observation times can then be defined as
the ratio between the luminosity produced by the same atmospheric shower
observed at the same Zenith angle but under the two different conditions. For
data obtained with the Whipple $10\U{m}$ telescope \cite{Cawley90,Finley01},
we define the luminosity of an event as the sum of the signals in all the
photomultiplier tubes (PMTs) that gave a significant contribution to the image
\cite{Reynolds93}. In order to effectively estimate the throughput factor we
use the fact that the cosmic ray spectrum is constant at the energies we
observe \cite{Gaisser90}, and therefore differences in the distribution of $Q$
obtained at the same Zenith angle with the same detector should only reflect
variations in light collection efficiency and gain of the
experiment. Practically, in the Whipple data analysis, we construct the
histogram of $Q$ obtained from the Zenith observations during a specific night.
This is then used as a reference for the other nights to be calibrated.  For
each of the other nights we construct the histogram of $F\times Q$, with $F$
being a test value for the relative gain between the night to be calibrated
and the reference night. We then adjust $F$ until the distribution best fits
the reference one. When applying this process one must take care to avoid the
regions of the $Q$ distribution where the effects of threshold and saturation
of the experiment become important.

\begin{figure}[t] 
\epsfig{file=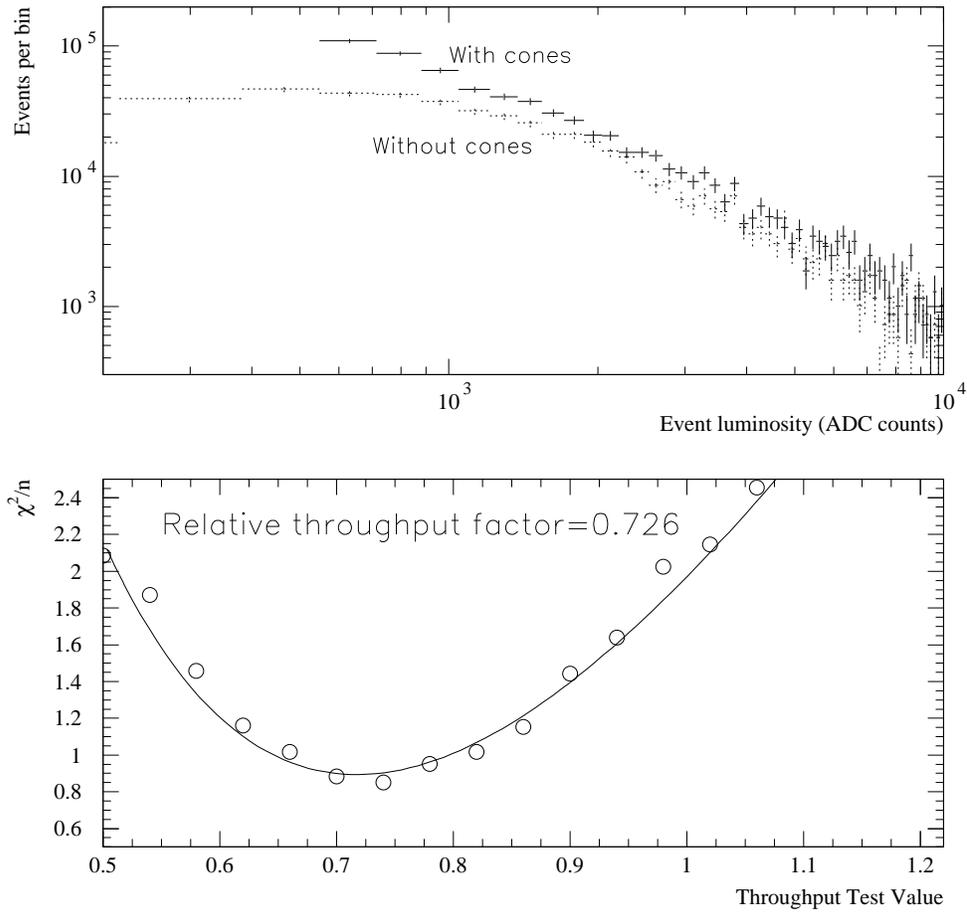,width=5.5in}
\vspace{10pt}
\caption{The event luminosity distributions obtained with and without light
collecting cones are shown at the top. The $\chi^2$ is calculated by comparing
the distribution obtained without the cones with the distribution obtained
with the cones (above a value of 1800 ADC counts) and rescaled by a test value
for the throughput factor $F$. The minimum occurs for $F=0.73\pm0.03$
indicating a $27\%$ contribution by the cones to the light collection
efficiency.}
\label{cones}
\end{figure}

\subsection{An Application: Calibration of the focal plane light collecting cones}
The Whipple telescope focal plane detector is equipped with light collecting
cones which reduce the dead space between the photomultiplier tube pixels. The actual light collection improvement due to the cones is given by a
combination of their optical properties and the shape and size of the main
optics as seen from the focal plane. This factor is usually estimated by
simulations based on measurements performed in the lab under simplified
conditions. Here we try to quantify this factor using the method described in
the previous section by taking some data toward the Zenith with and without the
cones installed. The comparison of the two sets of data is presented in
figure~\ref{cones}. The data with the cones are used as a reference and a
relative throughput factor is calculated for the data obtained without the
cones. From this we estimate that the cones are responsible for $27\%$ of the
Cherenkov light collection efficiency of the telescope. This result matches
very well with the previous estimates (F.Krennrich, private communication).

\subsection{Correlation with the sky quality}

\begin{figure}[] 
\epsfig{file=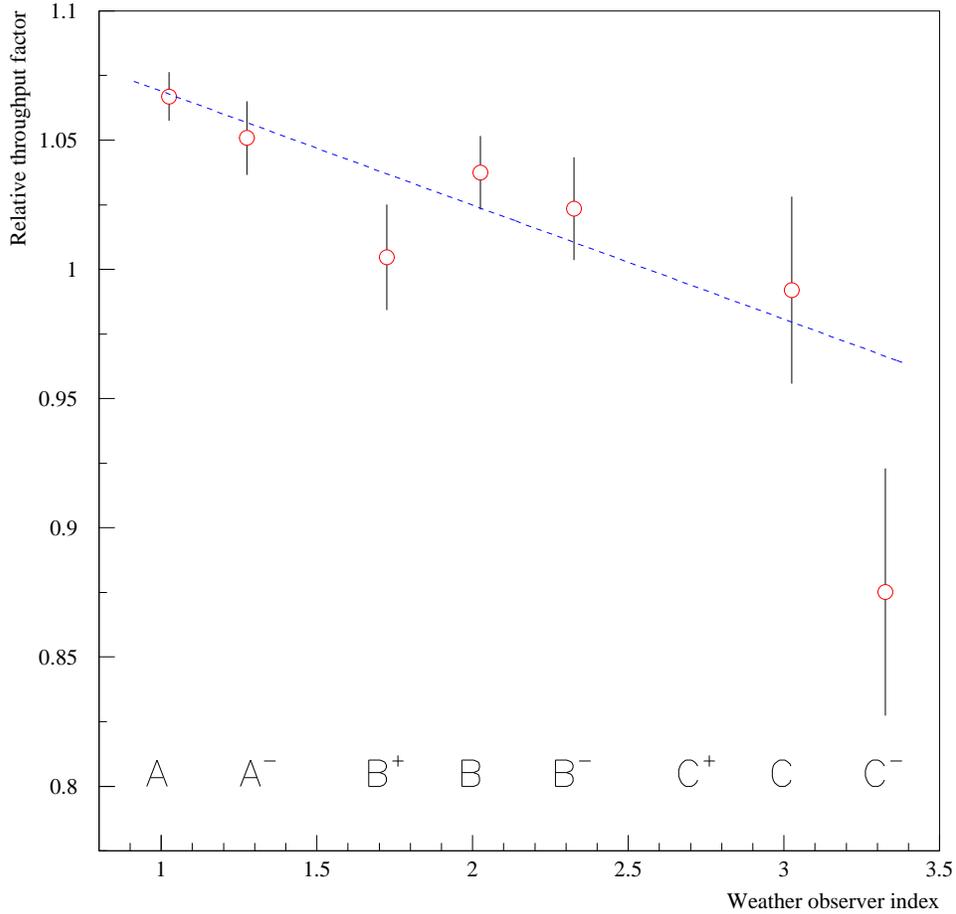,width=5.5in}
\vspace{10pt}
\caption{The relative throughput factor, $F$, measured at the Zenith as a
function of the observer's estimate of sky quality.}
\label{weather}
\end{figure}

Every night, the Whipple telescope operator records his estimate of the sky
quality on a qualitative scale ranging from $C^-$ to $A$. Data that were
obtained under sky qualities less than $B$ are often rejected. In
figure~\ref{weather} we show the average throughput factor as a function of
the observer's estimate of sky quality, with the vertical error bars
indicating the standard deviation. It appears that the throughput factor
increases as the sky quality improves but the plot also shows that some of the
data obtained under a $C$ sky could very well be used. Cirrus cloud occurs
typically at an altitude of $\sim10\U{km}$, with night to night variations of
$\pm3\U{km}$ \citep{Goldfarb00}. The majority of the Cherenkov light produced
by air showers is emitted near to the shower maximum (the altitude at which
the number of particles in the shower is greatest) which is $\sim10\U{km}$ for
showers initiated by a $500\U{GeV}$ $\gamma$-ray. It is possible, therefore,
that the observer can be influenced in his judgment of the weather conditions
by some high altitude cirrus clouds which could be responsible for an increase
in the sky background brightness without affecting the quantity of Cherenkov
light collected by the telescope. There will also be occasions when the
telescope was aimed between the clouds during the observation. Because of this,
the idea of calculating the throughput factor using a single test run with the
telescope pointing to the Zenith is invalid, as the results cannot be applied
to other observations on the same night. This is one of the motivations for
generalizing the method to any Zenith angle.

\section{Relative calibration at any Zenith angle}
\subsection{Generalization}

The relative throughput calibration method as described above already allows
us to correct for changes in the telescope which affect the light collection
efficiency, as well as helping with data selection. Nevertheless, it is based
on data obtained at a fixed Zenith angle, which cannot be strictly
contemporaneous with the astronomical observations of interest. Therefore it
can not be used with confidence to make corrections to the data. It is, in
principle, possible to apply the same method to compare data obtained at
different Zenith angles. The value of $F$ then results from differences in
atmospheric transparency as well as differences in the detection geometry
which affect both the energy threshold and the effective $\gamma$-ray
collection area. When observations are made at lower Zenith angles the
atmospheric showers produce a Cherenkov light pool which extends over a larger
area \citep{Sommers85}.

\begin{figure}[] 
\epsfig{file=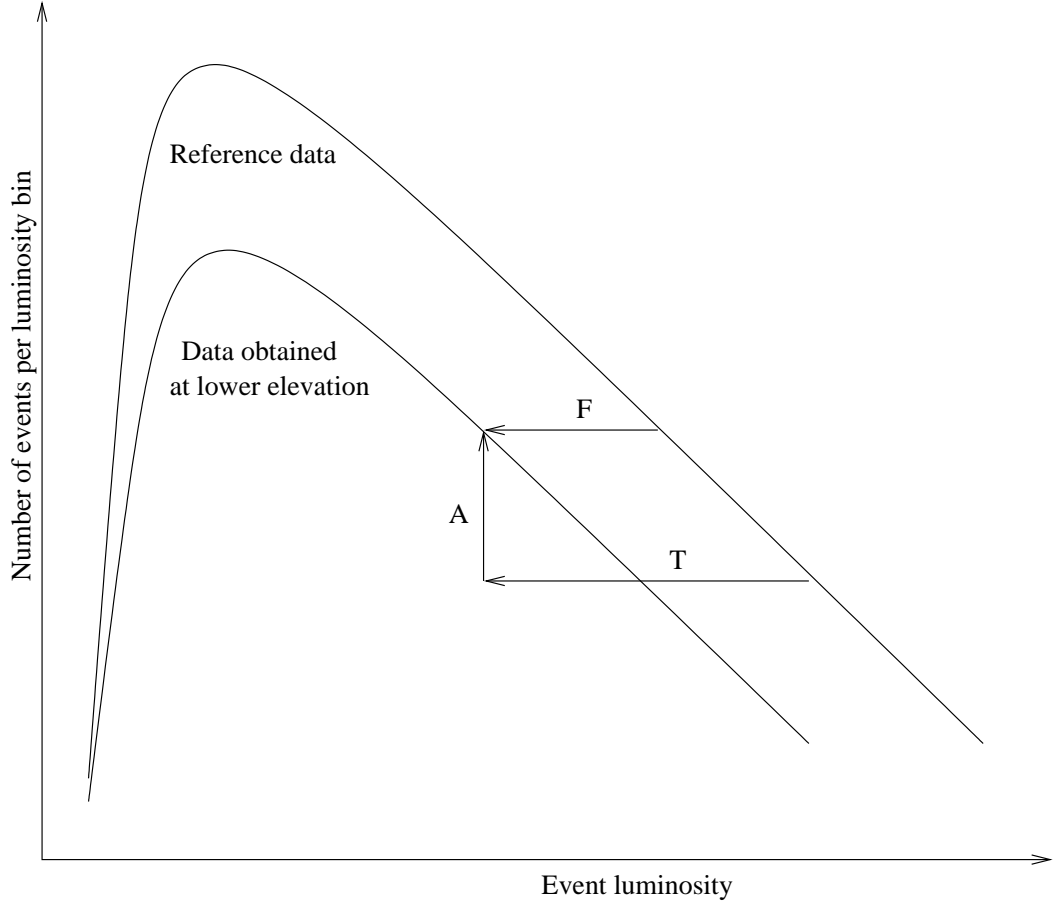,width=5.5in}
\vspace{10pt}
\caption{The general principle behind the throughput factor (both axes are in
log scale). See text for details}
\label{principle}
\end{figure}

Figure~\ref{principle} shows how the throughput factor $F$ corresponds to a
combination of two effects. Factor $T$ measures the horizontal shift in the
distribution caused by changes in the light collection efficiency which result
from changes in the instrument and atmospheric conditions as well as from
differences in Zenith angle. Factor $A$ measures the vertical shift (a change
in the number of showers observed with a given luminosity) caused by changes
in the effective $\gamma$-ray collection area due to different source Zenith
angles.  Only the factor $F$ can be directly measured from the event
luminosity distributions. As the radius of the Cherenkov light pool is defined
only by the atmospheric density profile and source Zenith angle, factors $T$
and $A$ should show the same Zenith angle dependence for both $\gamma$-ray and
cosmic-ray initiated showers as long as most of the Cherenkov light is emitted
from the core of the shower. This allows us to use a generalized throughput
factor in our analysis of $\gamma$-ray signals.

\subsection{Test and application of the method}
\begin{figure}[] 
\epsfig{file=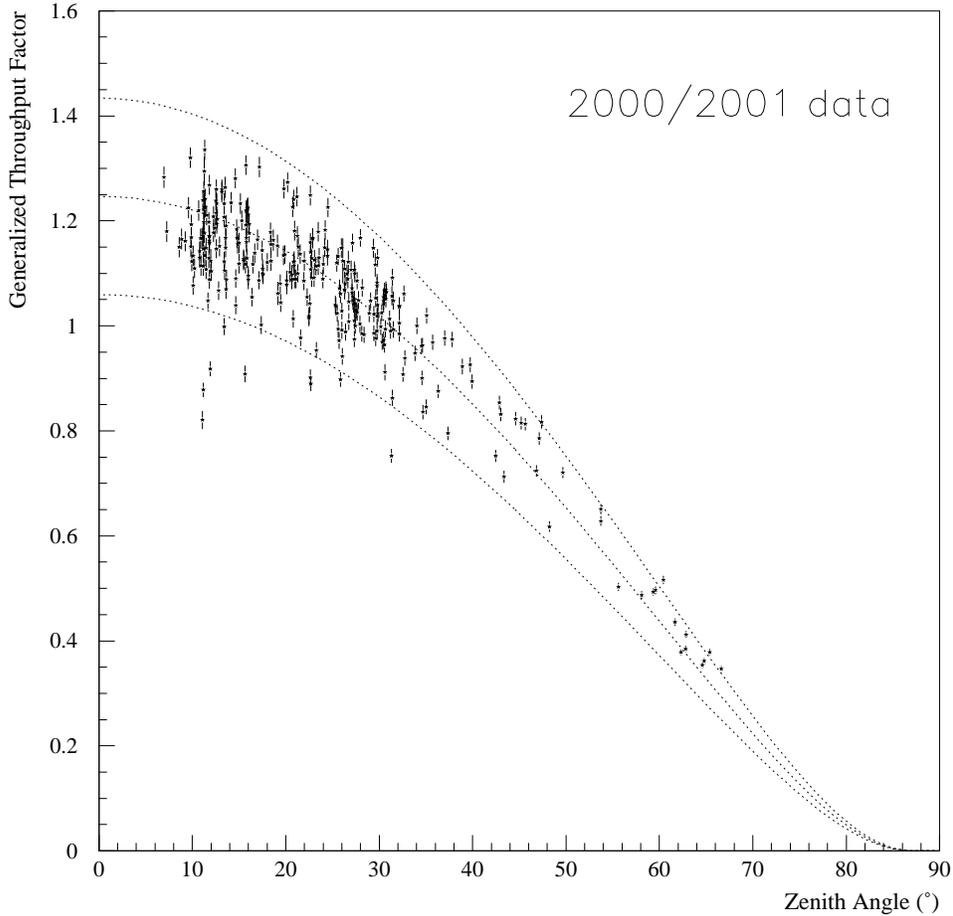,width=5.5in}
\vspace{10pt}
\caption{The throughput factor as a function of the distance from the
Zenith. Each point represents a 28 minute observation (with statistical
errors). The curves correspond to a simple isothermal model for the atmosphere
with three different values for atmospheric attenuation.}
\label{zendep}
\end{figure}

In figure~\ref{zendep} the throughput factor is shown as a function of
$\theta_z$, the distance from the Zenith. The reference data were taken at
$\theta_z\sim30^o$ from the Zenith and so $F$ is close to one at this point. It
can be shown that if the atmospheric density profile is assumed isothermal,
the area of the Cherenkov light pool is proportional to $\frac{1}{\cos^2
\theta_z}$ (see appendix). Using this, for a luminosity distribution 
of differential power law index $-\Gamma$, the throughput factor is 
expected to vary as

\begin{equation}
F \propto (\cos\theta_z)^{2(\frac{\Gamma-1}{\Gamma})}
\times \e^{-\frac{K}{\cos\theta_z}}
\end{equation}\label{zen1}

where the exponential term is used to describe the atmospheric attenuation of
Cherenkov light. For our observed $\Gamma=2.3$ we have

\begin{equation}
F \propto(\cos \theta_z)^{1.13} \times
\e^{-\frac{K}{\cos\theta_z}} 
\end{equation}\label{zen2}

This function gives the curves shown on figure~\ref{zendep} for three
different empirically derived values of $K$. Points falling near the upper
curve would correspond to data obtained under the best atmospheric conditions
while points on the lower curve correspond to data obtained under poorer
conditions. We can see on this figure that variations of $\pm20\%$ arise in
the event luminosity even when the observer estimated the sky quality to be
good (more than 90\% of these observations where graded as $A$ or $B$ weather
by the observer). Variations of this magnitude must be corrected for in order
to establish accurate $\gamma$-ray fluxes, particularly in the case of sources
with steep spectra.

In order to use the throughput value to correct the measured $\gamma$-ray
rate, we must verify that the $\gamma$-ray showers are affected by changes in
Zenith angle, instrument efficiency and atmospheric transparency in
approximately the same way as the background cosmic ray showers which are used
to derive the throughput factor. We do this by looking at the $\gamma$-ray
rate observed in the direction of the Crab Nebula: the Crab is the standard
candle of TeV $\gamma$-ray astronomy and dedicated studies (using selected
data taken close to Zenith and under good weather conditions) have shown its
emission at these energies to be constant over a timescale of years
\cite{Vacanti91}. If the throughput factor is applied correctly, the Crab
Nebula $\gamma$-ray rate after correction should remain stable within
statistical errors over all elevations and weather conditions

There are different ways in which one could apply the correction. One possible
method is to apply the throughput correction directly to the measured
Cherenkov photon yield in each PMT, prior to parameterization of the
images. The drawback here is that the signal to noise ratio will not remain
constant over the range of throughput factors and Zenith angles, and so the
efficiency of the $\gamma$-ray selection cuts will also change. In addition,
for observations where the correction factor is large, the hardware trigger
threshold will bias the number of images which pass the selection
cuts. Because of this, we prefer to apply the correction directly to the
measured rate. This is only strictly accurate if the spectrum of $\gamma$-rays
from the source follows a simple power law of known spectral index.

We try here to correct the $\gamma$-ray rate separately for the Zenith angle
and atmospheric transparency effects. The Zenith angle dependence of the
$\gamma$-ray rate is ideally calculated using Monte Carlo simulations; here we
use a simple analytical model which provides a good approximation. The
effective collection area, $A$, and threshold energy, $E_{th}$, are both
proportional to $\frac{1}{\cos^2\theta_z}$ and so the $\gamma$-ray rate
$\Phi\propto(\cos\theta_z)^{2(\alpha-1)}$ where $\alpha$ is the integral
$\gamma$-ray power law spectral index. For the Crab Nebula, $\alpha=1.5$
\citep{Hillas98} and so $\Phi\propto\cos\theta_z$. This can be used to correct
the measured $\gamma$-ray rate to the rate expected at a fixed Zenith angle;
we choose to calculate the corrected rate for a Zenith angle of $30^{\circ}$,
$\Phi_{30}$. 

To apply the throughput correction we first calculate the expected throughput
factor, $F_{exp}$, normalized to a Zenith angle of $30^{\circ}$ (because the
measured throughput factor $F_{meas}$ has been calculated with reference to an
observation taken at a Zenith angle of $30^{\circ}$) such that :

\begin{equation}
F_{exp} = (\frac{\cos \theta_z}{\cos 30^{\circ}})^{1.13}
\end{equation}\label{fexp}

This is equivalent to equation~2 but without atmospheric attenuation. The
effects of atmospheric attenuation are automatically incorporated in the
throughput correction , which we use to calculate the corrected rate as
follows:

\begin{equation}
\Phi_{corr}= \frac{\Phi_{30}}{(F_{meas}/F_{exp})^\alpha}
\end{equation}\label{tp}

\begin{figure*}[] 
\epsfig{file=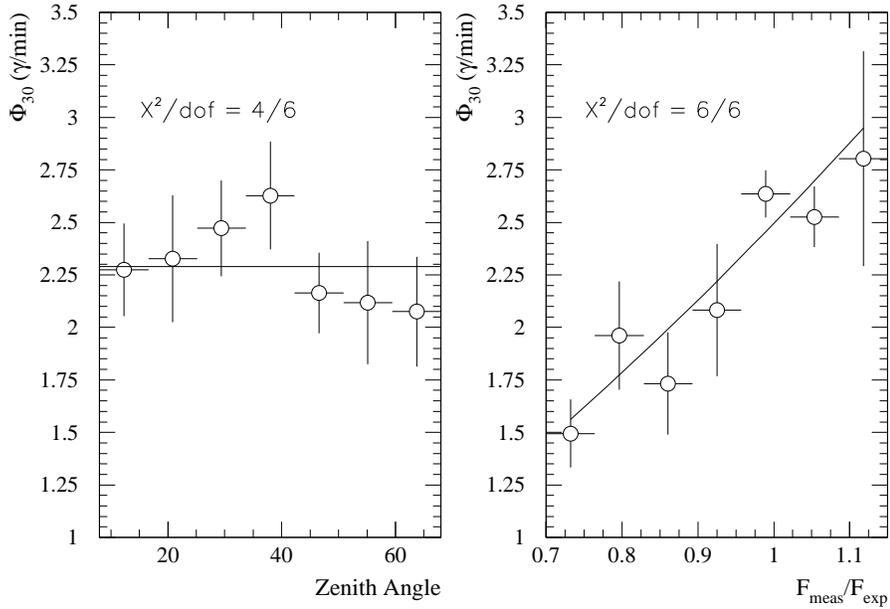,width=5.5in}
\vspace{10pt}
\caption{The averaged Crab nebula $\gamma$-ray rate after correction for the
Zenith angle as a function of Zenith angle (left) and $F_{meas}/F_{exp}$ (right).}
\label{crab_tp}
\end{figure*}

Figure~\ref{crab_tp} shows $\Phi_{30}$ as a function of Zenith angle and of
$F_{meas}/F_{exp}$. The $\gamma$-ray rate is constant with Zenith angle
after the Zenith angle correction, while there is clearly still a correlation
with the throughput correction which is well fit by a power law of index
$\alpha=1.5$, as expected for the Crab.

\begin{figure*}[] 
\epsfig{file=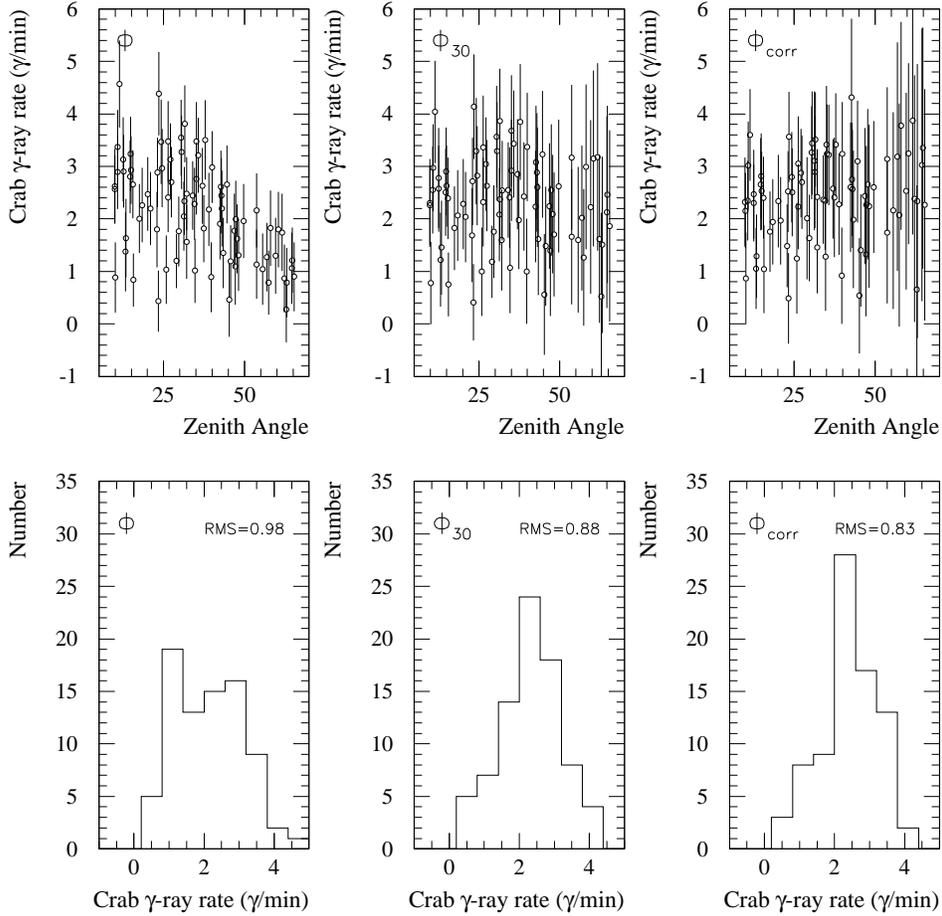,width=5.5in}
\vspace{10pt}
\caption{The effect of the elevation and throughput corrections to the Crab
Nebula $\gamma$-ray rate. The upper plots show the rate as a function of the
Zenith angle. Each point represents a 28 minute observation (with statistical
errors) showing the uncorrected rate ($\Phi$), the rate corrected to a
fixed Zenith angle ($\Phi_{30}$) and the rate corrected for Zenith angle and
throughput ($\Phi_{corr}$). The lower plots are histograms showing the
distribution of the three rates.}
\label{crab_el}
\end{figure*}

Figure~\ref{crab_el} shows the reduction in the width of the rate
distributions at each stage of the correction. We note, however, that the
width of this distribution is not the best measure of the effectiveness of the
correction. Observations at large Zenith angles often result in a measurement
which is not statistically significant and so the width of the distribution
for these runs will be dominated by statistical fluctuations. Most of these
Crab data were taken under good weather conditions; the throughput correction
will play an even stronger role when trying to analyse data taken under poorer
conditions.

\subsection{Application to Mrk421 Variability Studies}
Markarian 421 (Mrk421) is a bright TeV $\gamma$-ray source which has been well
studied and is known to be extremely variable \cite{Buckley96,Aharonian99,Piron01}.
During the 2000/2001 observing season this source was in the most active state
yet observed \cite{Holder01}, with an average TeV $\gamma$-ray flux of 1.5
times the steady flux from the Crab Nebula.  Measurements in the $2-12\U{keV}$
X-ray region were also made by the All Sky Monitor (ASM) on board the Rossi
X-Ray Timing Explorer (RXTE).  Multiwavelength studies of the variable
emission from AGN are extremely important to our understanding of the nature
of the particles and acceleration mechanisms in jets.  In the 2000/2001
Whipple observations of Mrk421, data were taken whenever possible, including
at low Zenith angles and during poor weather, so as to provide the best
possible sampling of the $\gamma$-ray light curve.  The throughput factor
provides us with a method to treat these data in a consistent fashion.

\begin{figure}[] 
\epsfig{file=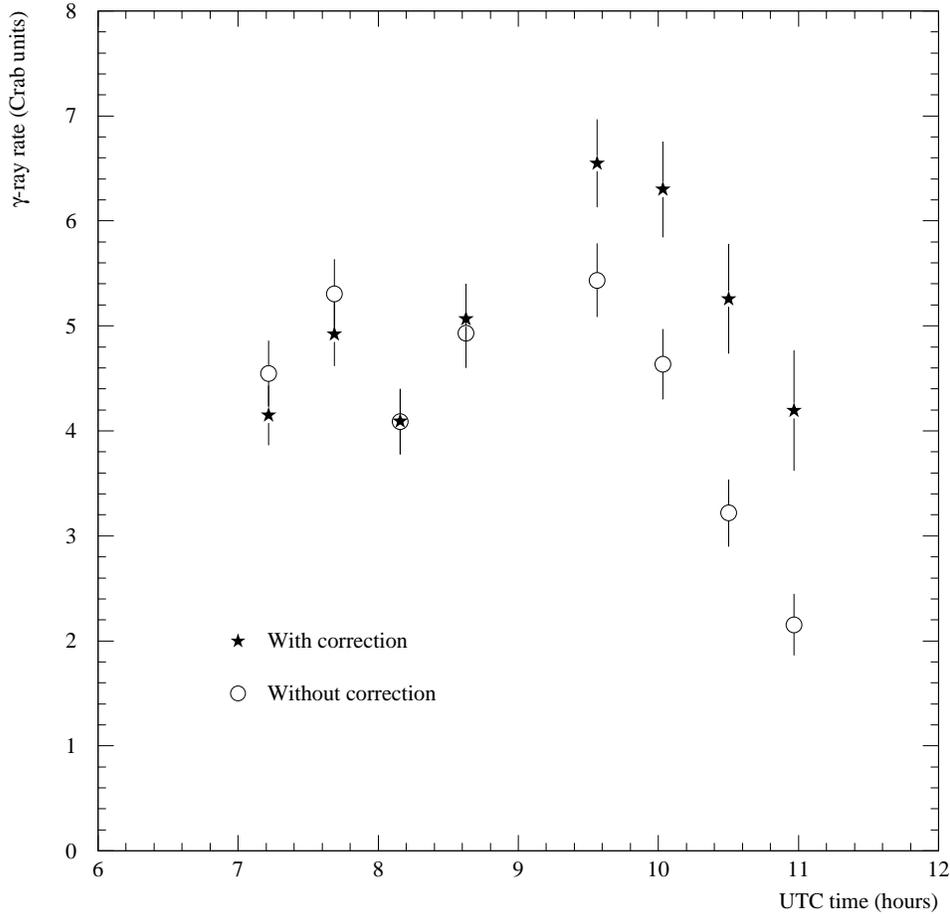,width=5.5in}
\vspace{10pt}
\caption{The light curve of Mkn421 for the observations of March $27^{th}$
2001 with and without the correction for elevation and throughput factor
being applied.}
\label{flare}
\end{figure}

Observations of Mrk421 for the night of March $27^{th}$ 2001 are shown in
figure~\ref{flare}. The source was in a very high state and so we observed
continuously from a source Zenith angle of $12^{\circ}$ down to $60^{\circ}$.
The difference between the results before and after throughput factor
correction illustrates the importance of this correction when studying the
detailed structure of flares.

\begin{figure*}[] 
\epsfig{file=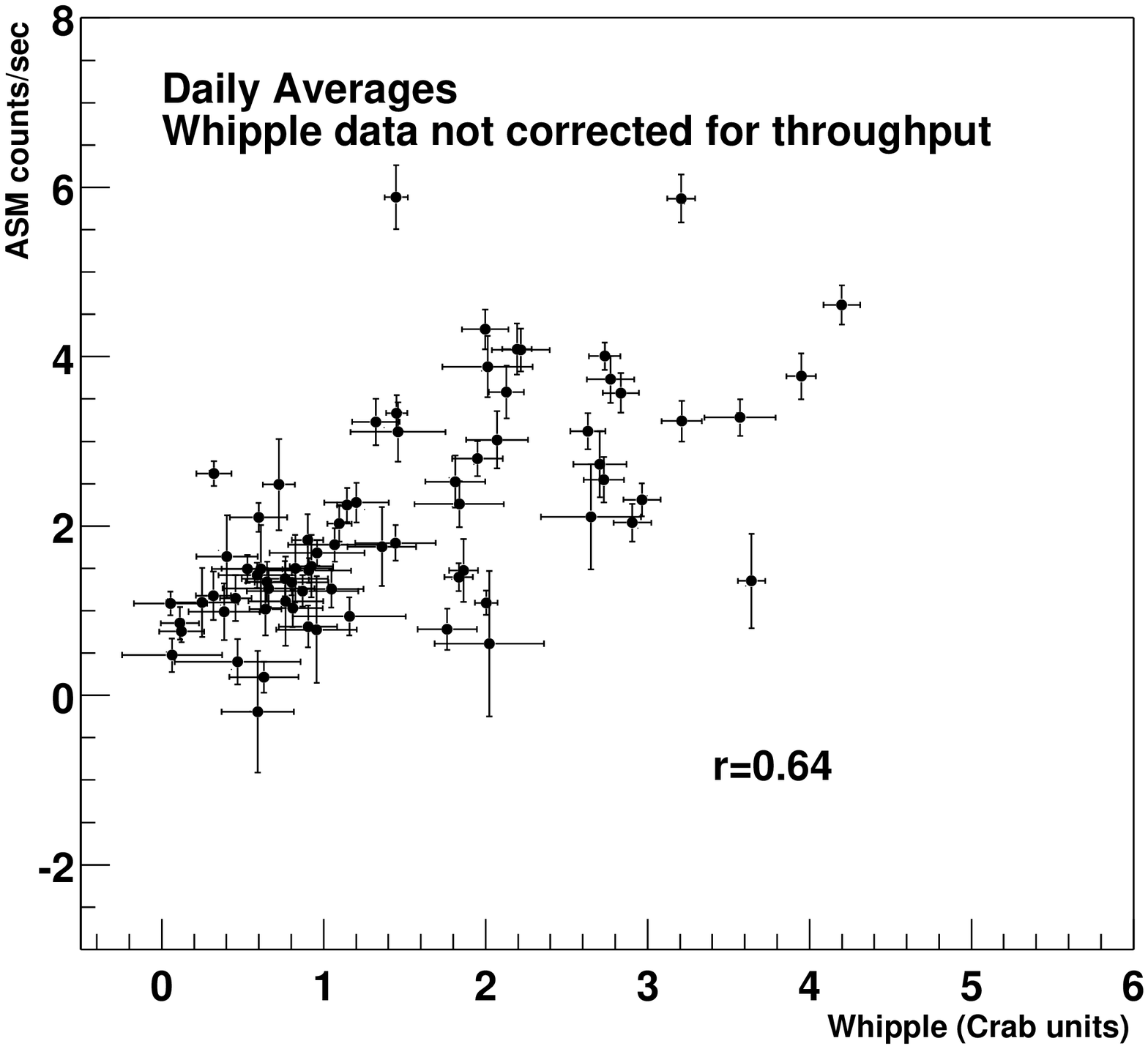,width=2.75in}\epsfig{file=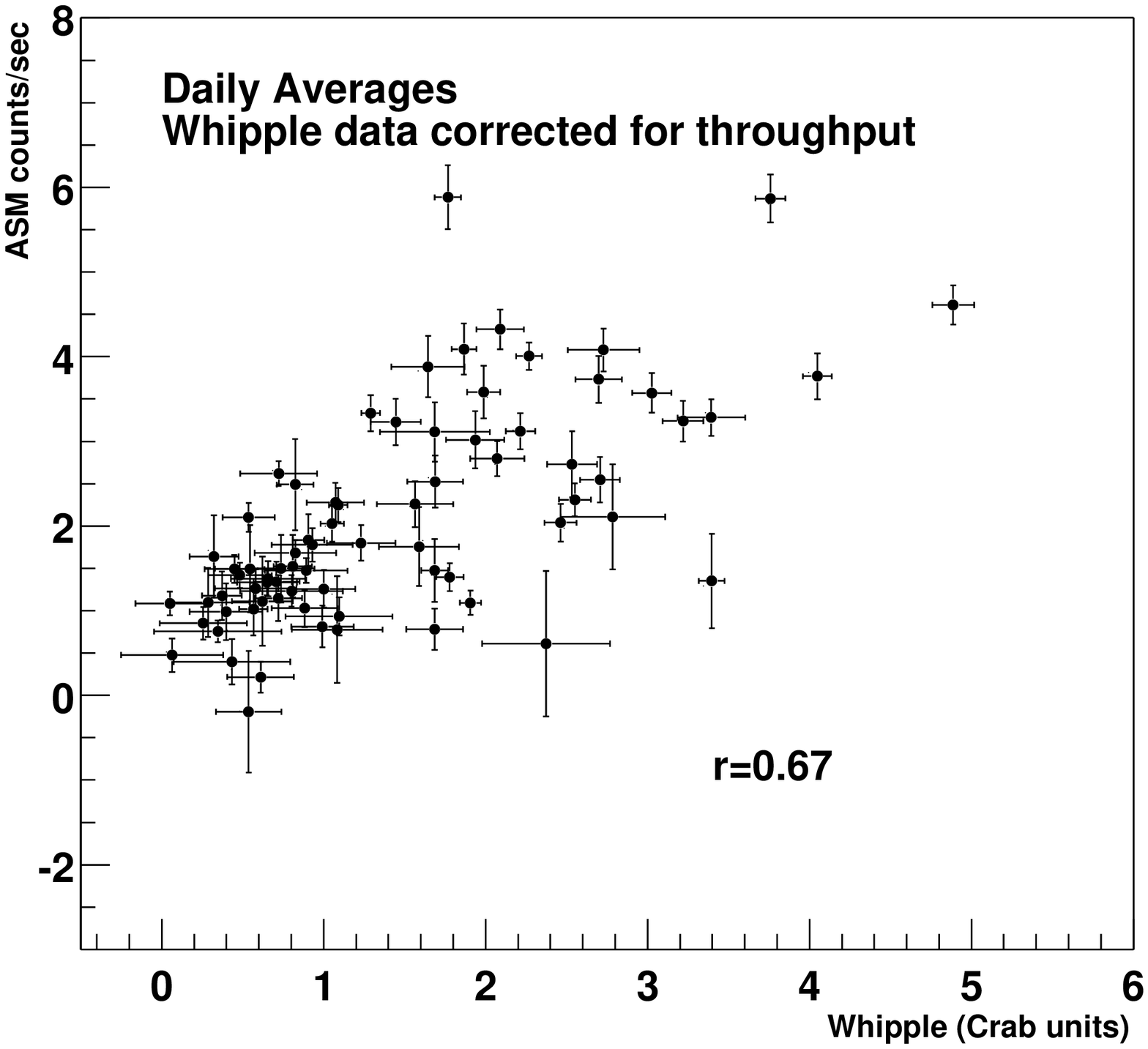,width=2.75in}
\vspace{10pt}
\caption{Correlation between daily averaged ASM quicklook flux and $\gamma$-ray flux for Mrk421. In the right-hand plot the throughput correction has been applied, in the left-hand plot it has not.}
\label{mrk421}
\end{figure*}

Figure~\ref{mrk421} shows the correlation between X-ray and $\gamma$-ray
measurements both with and without a throughput correction applied to the
$\gamma$-ray data.  The correlation, as defined by the linear correlation
coefficient, $r$, improves slightly when the throughput correction is
applied. The correlation is not perfect; the remaining scatter may be due to
the fact that the measurements are daily average fluxes, where the X-ray and
$\gamma$-ray observations which have been used to calculate the averages are
not exactly contemporaneous. Alternatively, it may be due to the details of
the $\gamma$-ray production mechanism in the source. Clearly though, the
magnitude of the effect of the throughput correction on the measured
$\gamma$-ray flux shows the importance of the correction in this type of
study.

\section{Relative Calibration of a Telescope Array: CELESTE}

In this section we consider a different use of the cosmic ray
background. Rather than attempting to correct for temporal changes in the
efficiency of a single telescope we are interested in calculating a relative
calibration between the different elements of an array.

The CELESTE experiment \cite{proposal,NIM}, situated in the French Pyrenees,
uses forty movable $54\UU{m}{2}$ mirrors (known as heliostats) of a former
solar electrical plant to reflect Cherenkov light from air showers to a
detector package at the top of a $100\U{m}$ tall tower. The detector package
consists of a single photomultiplier tube (PMT) equipped with a fast
analog-to-digital converter (FADC) for each heliostat, providing a measure of
the arrival time and photon density of the Cherenkov light at ground
level. This type of Cherenkov wavefront sampling experiment provides a massive
total mirror area which allows us to reach an energy threshold of $60\U{GeV}$
\citep{Denaurois02}. The fundamental problem still exists however; there is no
test beam with which to calibrate the experiment. Furthermore, in the case of
a heliostat array it is necessary to calculate a relative calibration of the
different heliostats which, for a solar plant experiment like CELESTE, will
vary as the source position is tracked across the sky and the mirrors present
a changing area of reflecting surface to the detector.

For each event which triggers the experiment, the luminosity $Q$ for each
heliostat is simply given by the charge measured by the single PMT. The
histograms of $Q$ can then be used to calculate a throughput factor for each
heliostat in a similar fashion to that described above for the Whipple
telescope. The difference in this case is that the throughput factor describes
the relative gains of the forty heliostats. In fact, the throughput
calculation has been made in a simplified way for CELESTE by simply
integrating the $Q$ histograms and measuring the values of $Q$ at two constant
fraction levels ($5\%$ and $30\%$) of the total number of events. These
fractions were chosen so as to be distant from the regions of the histograms
affected by saturation of the ADCs at the higher end and by the trigger
conditions at the lower. We then normalize the difference in $Q$ at these two
levels to provide the throughput measurement for each heliostat. If the
distribution of $Q$ is a perfect power law this method should produce
identical results to the $\chi^2$ minimization used for the Whipple telescope
data. We have compared the two methods using Whipple data taken over a range
of elevations and find the results to be consistent to within $\sim10\%$. The
results of this section illustrate relative changes in the heliostat response
and so are not strongly affected by this discrepency.

\begin{figure*}[] 
\begin{center}
\epsfig{file=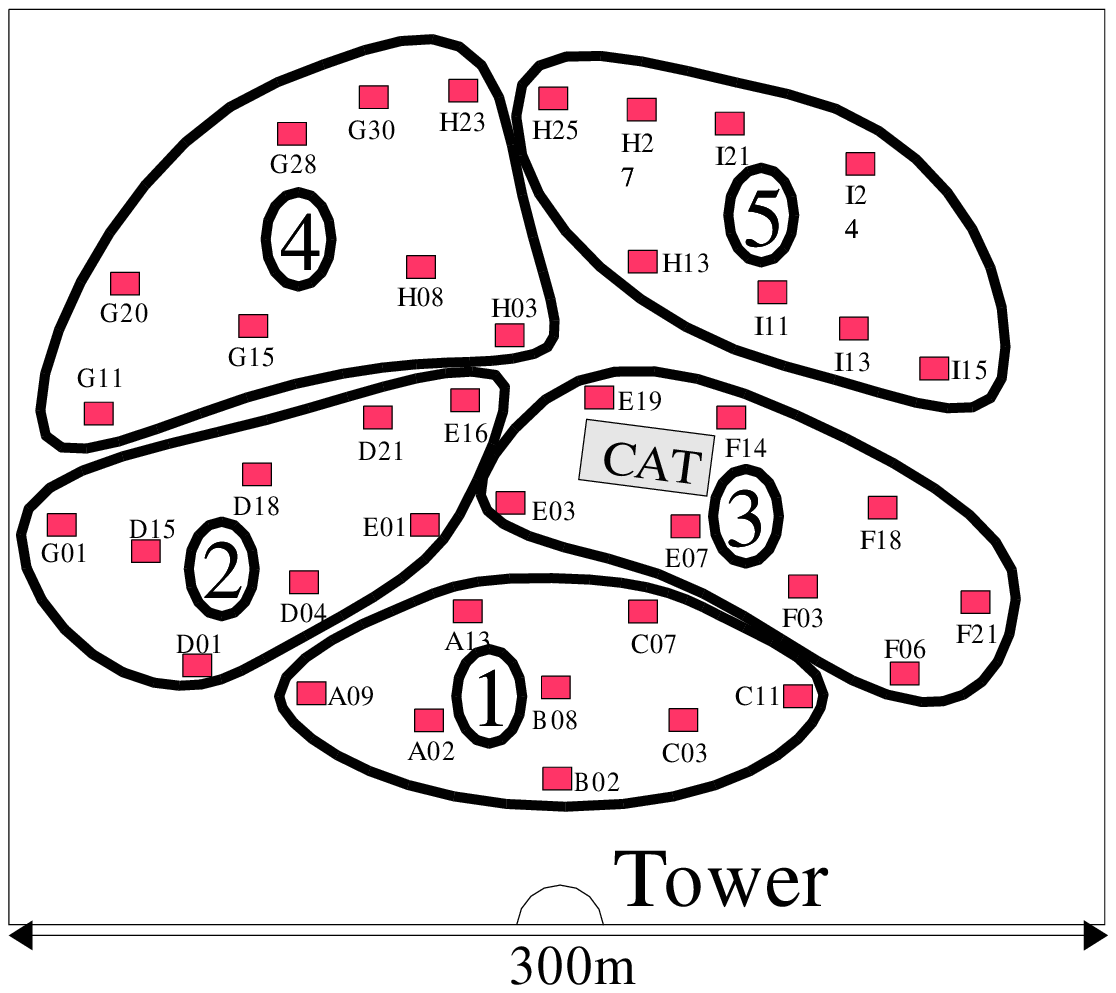,width=4in,height=3in}
\epsfig{file=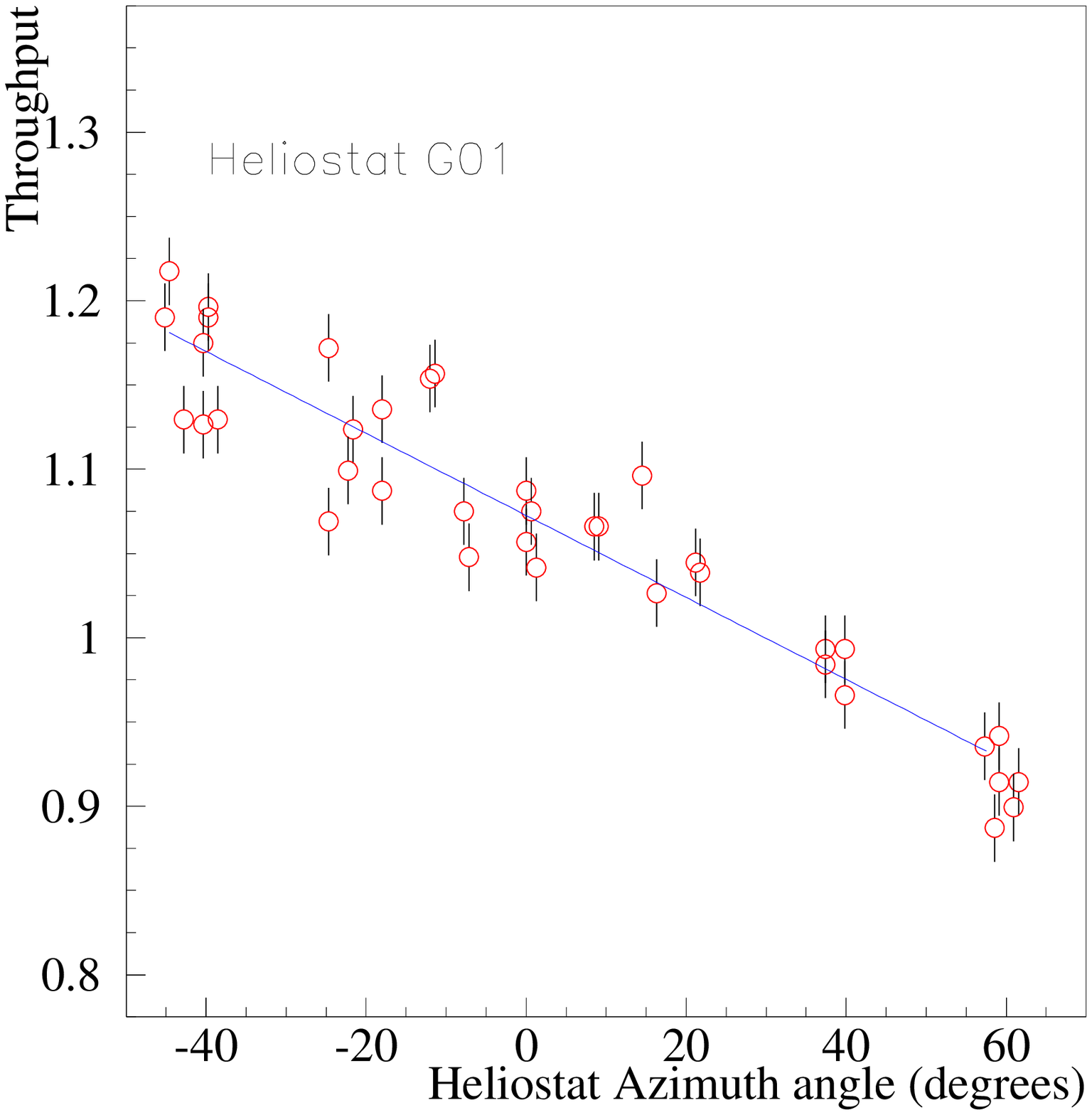,width=2in,height=2in}\epsfig{file=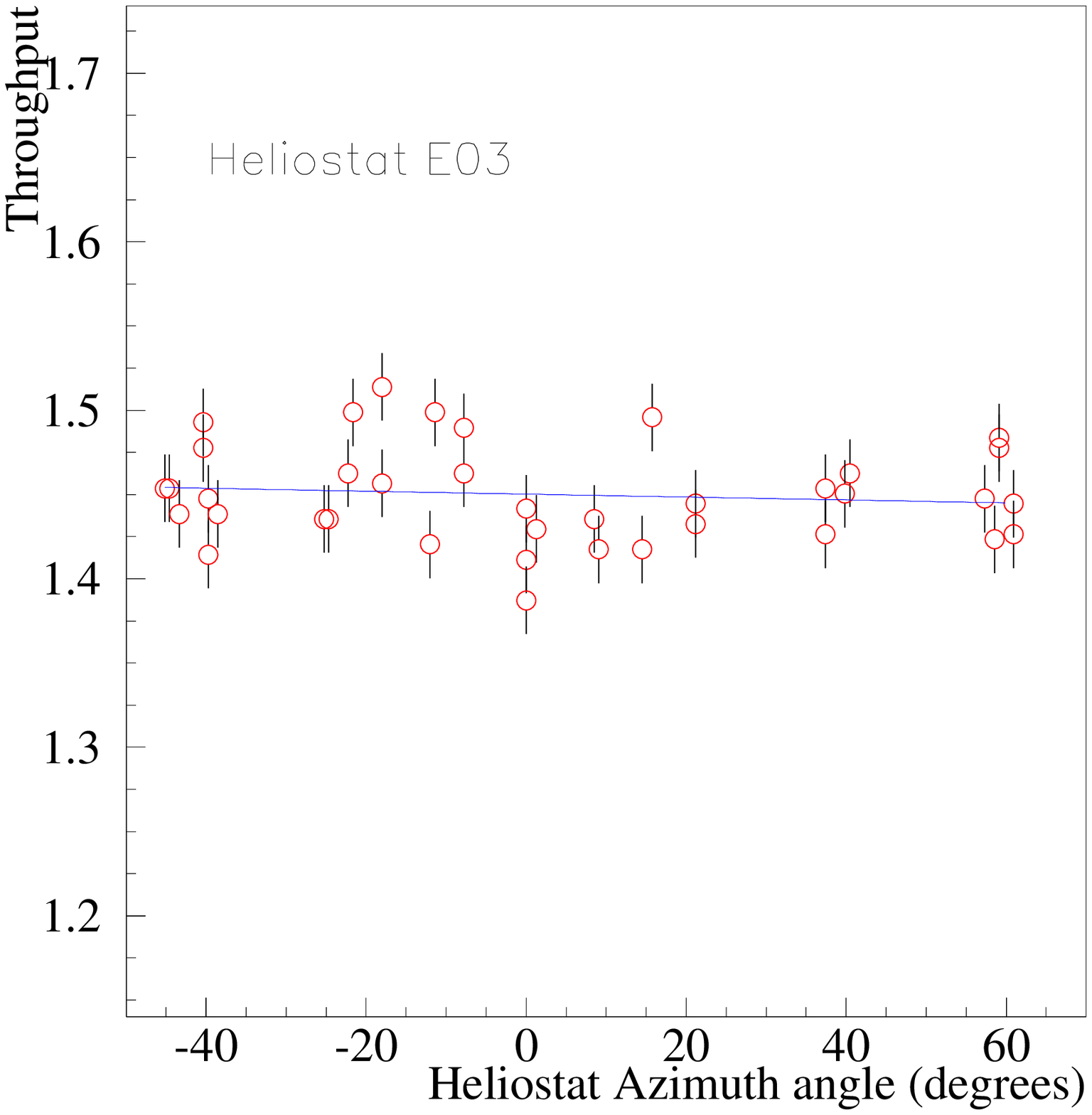,width=2in,height=2in}\epsfig{file=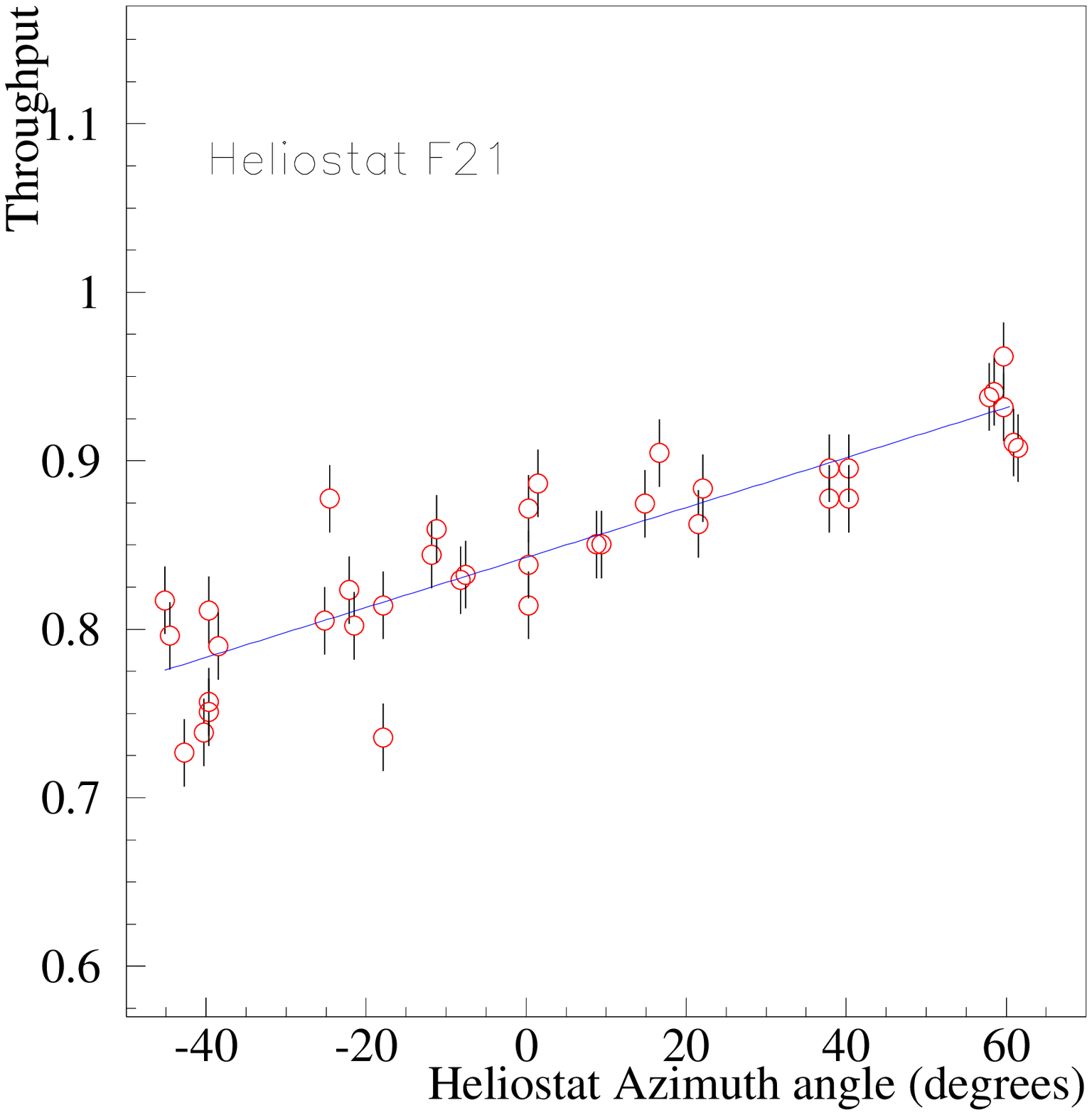,width=2in,height=2in}
\epsfig{file=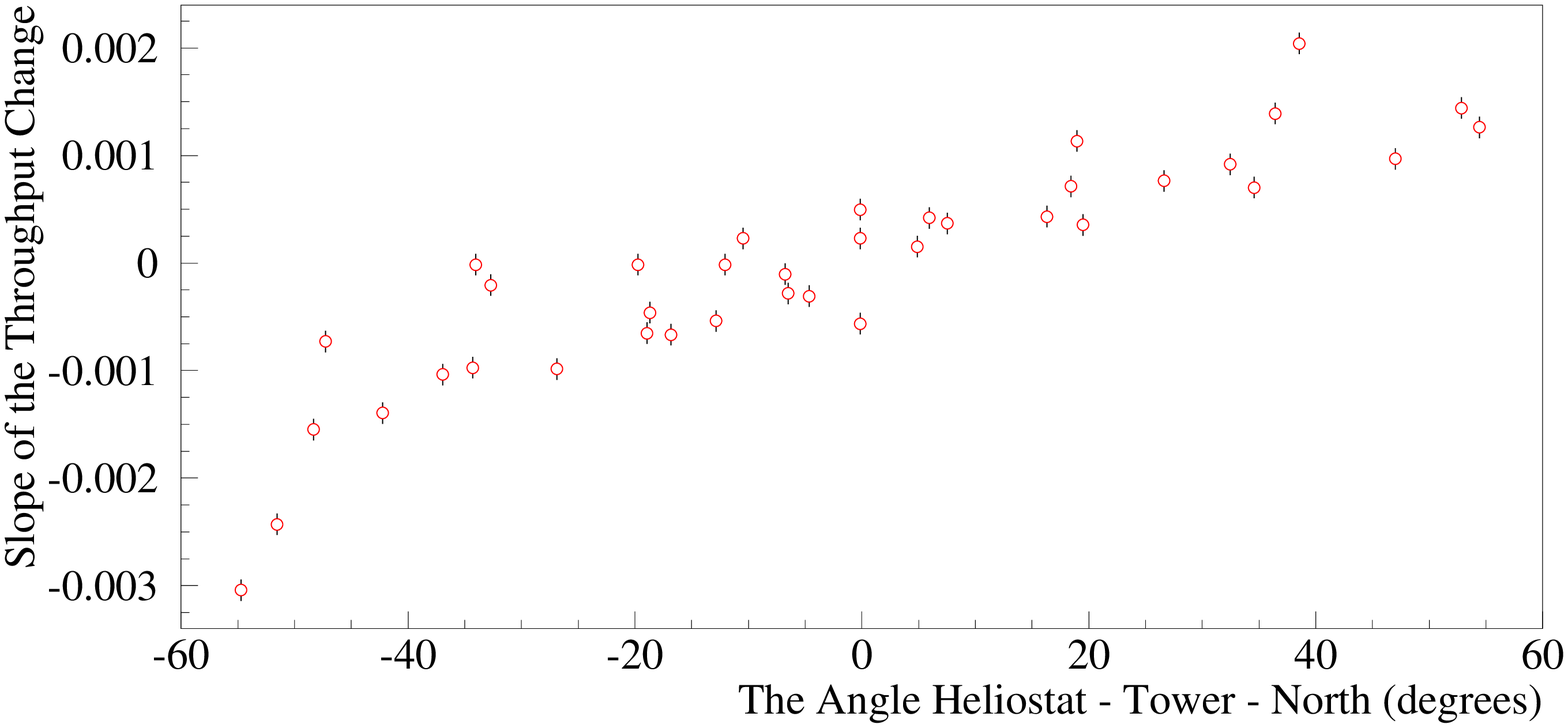,width=6in,height=3.5in}
\end{center}
\vspace{10pt}
\caption{The throughput variation with heliostat Azimuth angle as a function
of the heliostat position. See text for details.}
\label{celeste}
\end{figure*}

An example of the use of this measurement is shown in
figure~\ref{celeste}. The upper plot shows the layout of the CELESTE heliostat
array and the tower which houses the detector package. Also shown are the five
groups of heliostats whose signals are summed and used to trigger the
experiment and the position of the Cherenkov imaging experiment, CAT.  The
three central plots show the throughput measurements for three heliostats at
different positions in the heliostat array as a function of the Azimuth angle
of the source in the sky. As the source moves across the sky, the heliostat
efficiency changes, and this change will be greatest for heliostats which
reflect the Cherenkov light through the largest angle to the detector in the
tower. The three plots correspond to heliostats located on the far left
(west), centrally and on the far right (east) of the array. The slope of the
measured change in throughput with Azimuth angle reflects these
positions. This is illustrated further in the lower plot which shows the slope
of the throughput change with Azimuth angle as a function of the heliostat's
angular position in the array defined by the angle heliostat - tower -
north. A heliostat due north of the tower has an angular position of
$0^{\circ}$, while those to the east and west of the tower have positive and
negative angular positions, respectively.

The measured change in throughput with source position has been used to verify
the simulation of the rather complex telescope optics with good results
\citep{Denaurois00}. Also, the heliostat throughput factors, calculated with
the heliostats observing a point due south of the experiment, have been used
to calculate an adjustment to the high voltage supply of each PMT such that
the relative gain after the adjustment is approximately the same for each
heliostat/PMT pair. This correction smooths out the largest differences
between the different groups of heliostats which are used to trigger the
experiment and prevents the experiment energy threshold being defined by a few
very sensitive detectors. The throughput factors have also been used to
identify heliostats with erratic behaviour; for example, heliostats with large
tracking errors, which can then be repaired.

\section{Conclusion}
We have shown that cosmic ray background events observed at fixed Zenith angle
can be used to establish a relative calibration for a single atmospheric
Cherenkov imaging telescope in order to account for the many unavoidable
temporal changes in light collection efficiency, gain and, most importantly,
atmospheric conditions. Generalizing the method, we have shown that it can be
used for the relative calibration of data obtained at different Zenith angles,
taking into account both the geometrical effects due to Zenith angle and the
variations in atmospheric conditions.

This calibration method can be used to introduce corrections at various
levels. At the most basic level, it can be used to select which data were
taken under good conditions. We have also shown that it can be used to rescale
the measured $\gamma$-ray fluxes in order to make observations taken under
different conditions more comparable. One method of estimating the background
due to cosmic rays for $\gamma$-ray observations taken without dedicated
background control observations is to choose archival background observations
taken under conditions as similar as possible to the source observation being
considered. The throughput factor can be used as one of the criteria to judge
which background runs are most suitable \citep{Horan02}.

The application of the throughput calibration to CELESTE observations shows
how useful it may be when considering telescope arrays. The next generation of
Cherenkov imaging telescopes are currently being developed; the VERITAS
\citep{VERITAS}, HESS \citep{HESS} and CANGAROO III \citep{CANGAROO} projects
all involve using multiple telescopes on the same site. Inter-calibration of
these telescopes will be difficult without a dedicated test beam. For VERITAS,
simulations indicate that an energy resolution of $15\%$ should be possible;
in practice, this will require a relative calibration accurate to $<15\%$. The
throughput method described here may well prove to be the best solution

\section{Acknowledgements}
We gratefully acknowledge the VERITAS and CELESTE collaborations for the use
of their data. We would also like to thank Philippe Bruel for his help in
developing the CELESTE calibration.

\appendix
\section{Appendix}

\begin{figure}[] 
\epsfig{file=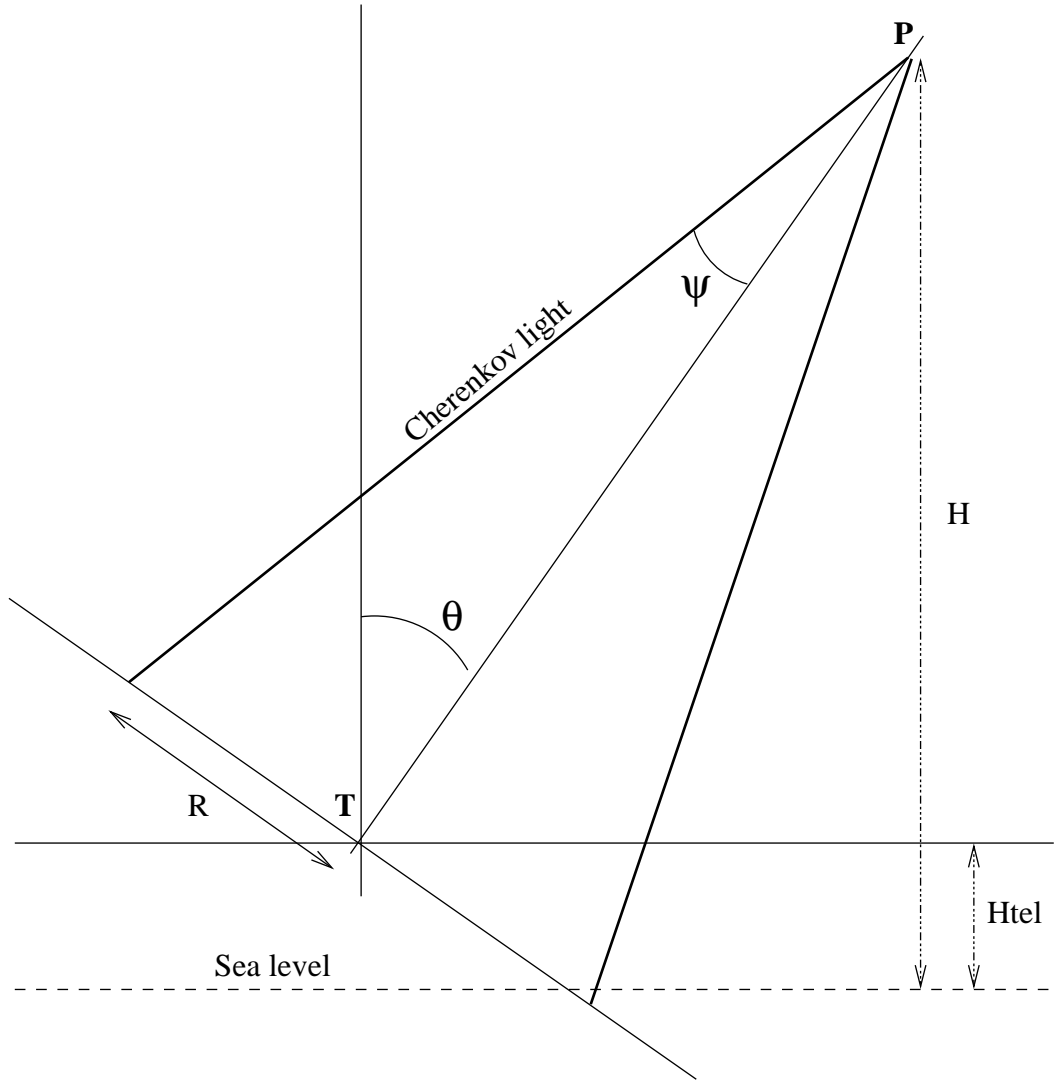,width=5.5in}
\vspace{10pt}
\caption{$\psi$ is the Cherenkov angle. $\theta$ is the angle between the
particle arrival direction and the vertical. $R$ is the radius of the circle
drawn by the Cherenkov light emitted at $P$ on the plane perpendicular to the
arrival direction at $T$, the position of the telescope. $H_{tel}$ is the
telescope altitude above sea level.}
\label{lightpool}
\end{figure}

Figure~\ref{lightpool} shows schematically the emission of Cherenkov light in
the atmosphere. We can see that $R={{H-H_{tel}}\over{\cos\theta}}\tan\psi$.
The Cherenkov angle $\psi$ is given by $\cos\psi=1/n$ where $n$ is the
atmospheric refraction index at altitude H. In a standard isothermal
atmosphere $n=1+273\times 10^{-6}e^{-H/8.5}$ where H is expressed in
km. Using the first order small angles approximation for $\psi$ one finds
$R={{H-H_{tel}}\over{\cos\theta}}\sqrt{(546\times 10^{-6}e^{-H/8.5})}$.

For small values of $H$, $R$ increases with $H$ while for large values of $H$,
 $R$ decreases with $H$. Therefore $R$ must take a maximum of value $R_{max}$ 
which corresponds to the rim of the Cherenkov light pool. By solving 
${{dR}\over{dH}}=0$ for R one finds:
$R_{max}={{17}\over{\cos\theta}}\sqrt{546\times10^{-6}e^{-(17+H_{tel})/8.5}}$
which, for $H_{tel}=2km$, gives $R_{max}\sim{{130m}\over{\cos\theta}}$.
From this it results that the effective collection area should scale as
${{1}\over{\cos^2\theta}}$.

When we consider cosmic rays which are incident at an angle to the vertical
the air shower develops in less dense atmosphere, where the Cherenkov emission
per unit of track length is lower. The total track length is longer by an
amount which compensates for this and consequently the total quantity of
Cherenkov light produced by the shower does not depend on the Zenith angle
$\theta$. However, as the Cherenkov light pool on the ground extends over a
larger radius, the light is more diluted and $Q$, the measured luminosity of a
shower, scales as ${{\cos^2\theta}}$.

In the generalized throughput calculation we compare luminosity distributions
obtained under different elevations. The content of a specific luminosity bin
will be affected by a factor ${{1}\over{\cos^2\theta}}$, corresponding to the
change in collection area, and by a factor $\cos^{2\Gamma}\theta$,
corresponding to the event luminosity scaled for a power law luminosity
distribution of spectral index $\Gamma=2.3$ as measured in the case of the
Whipple telescope. By combining those two factors raised to the ${1/\Gamma}$,
one expects the throughput factor to scale as $F \propto
(\cos\theta_z)^{2(\frac{\Gamma-1}{\Gamma})}$.

\end{document}